\documentclass[aps,prb,reprint,amssymb,amsmath,superscriptaddress]{revtex4-1}

\usepackage{graphicx}
\usepackage{epstopdf}
\usepackage[english]{babel}
\bibpunct{[}{]}{,}{n}{}{}

\begin{document}

\title{Zeeman splitting of conduction band in HgTe quantum wells near the Dirac point}

\author{G.\,M.~Minkov}

\affiliation{Institute of Natural Sciences, Ural Federal University,
620002 Ekaterinburg, Russia}
%
%\author{A.\,V.~Germanenko}

\author{O.\,E.~Rut}
\affiliation{Institute of Natural Sciences, Ural Federal University,
620002 Ekaterinburg, Russia}

\author{A.\,A.~Sherstobitov}

\affiliation{Institute of Natural Sciences, Ural Federal University,
620002 Ekaterinburg, Russia}

\affiliation{M.~N.~Miheev Institute of Metal Physics of Ural Branch of
Russian Academy of Sciences, 620137 Ekaterinburg, Russia}

%\author{M.\,O.~Nestoklon}

%\affiliation{Ioffe Physical-Technical Institute, Russian Academy of Sciences, 194021 St. Petersburg, Russia}

\author{S.\,A.~Dvoretski}

\affiliation{Institute of Semiconductor Physics RAS, 630090
Novosibirsk, Russia}

\author{N.\,N.~Mikhailov}

\affiliation{Institute of Semiconductor Physics RAS, 630090
Novosibirsk, Russia}
\affiliation{Novosibirsk State University, Novosibirsk 630090, Russia}

\date{\today}

\begin{abstract}
The Zeeman splitting of the conduction band in the HgTe quantum wells both with normal and inverted spectrum has been studied experimentally in a wide electron density range. The simultaneous analysis of the SdH oscillations in low magnetic fields at different tilt angles and of the shape of the oscillations in moderate magnetic fields gives a possibility to find the ratio of the Zeeman splitting to the orbital one and anisotropy of $\textsl{g}$-factor. It is shown that the ratios of the Zeeman splitting to the orbital one are close to each other for both types of  structures, with a normal and inverted spectrum and they are close enough to the values calculated within \emph{kP} method. In contrast, the values of $\textsl{g}$-factor anisotropy in the  structures with normal and inverted spectra is strongly different and for both cases differs significantly from the calculated ones. We believe that such disagreement
with calculations is a result of the interface inversion asymmetry in the HgTe quantum well, which is not taken into account in the \emph{kP} calculations.
\end{abstract}


\maketitle

\section{Introduction}
\label{sec:intr}

\begin{figure}
\includegraphics[width=1\linewidth,clip=true]{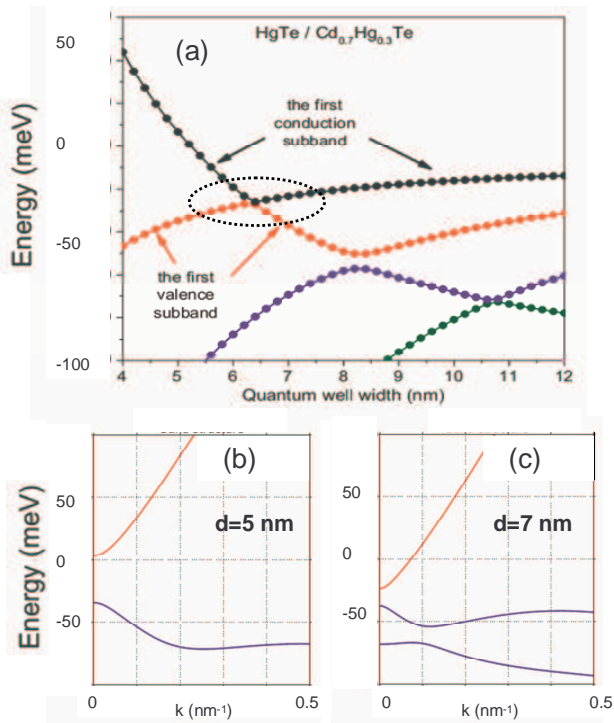}
\caption{(Color online) (a) -- The quantum well width dependence of the subband energies at $k=0$.  The dependences $E(k)$ of the conduction and valence bands for $d<d_c$ (b) and $d>d_c$ (c). The marked area in (a) shows the range of quantum well widths under study.}
\label{F1}
\end{figure}

A HgTe/CdTe quantum well is a system where the Dirac fermions appear only in a single valley, at the $\Gamma$ point of the Brillouin zone, unlike graphene where there are two valleys of the Dirac fermions with a strong inter-valley scattering.
The energies of spatially quantized sub-bands at the quasimomentum $k=0$ and energy spectrum $E(k)$ for different  widths of the quantum well  ($d$) were calculated within \emph{kP} method in numerous papers \cite{Gerchikov90,Zhang01,Novik05,Bernevig06,ZholudevPhD}. As seen from Fig.~\ref{F1},  various types of energy spectrum are realized upon increasing the HgTe quantum-well width; namely,  ``normal'', when $d$ is less than a critical width $d_c\simeq 6.3$~nm, Dirac-like at small quasimomenta  for $d = d_c$, inverted when $d>d_c$, and finally, semimetallic when $d>14-16$~nm. To interpret  experimental data, these calculations of the energy spectrum  are used practically always. They well describe the width dependence of the energies of  both  electron and hole  subbands at $k=0$ and the energy dependence of the electron effective mass ($m_e$).

However, quite a lot of differences between the experimental data and the results of these calculations on the energy spectrum of the carriers  have been accumulated to date. First of all, they refer to the spectrum of the valence band. The  hole effective mass ($m_h$) at $d\approx 20$~nm within the wide hole density range $p=(1-4)\times 10^{11}$~cm$^{-2}$ is substantially less than the  calculated one: $m_h\simeq (0.15-0.3)m_0$ \cite{Kozlov11,Minkov13} instead of $(0.5-0.6)m_0$ \cite{ZholudevPhD}. The top of the valence band in the nominally symmetric structures with $d\approx d_c$ ($d=5.5-7$~nm) was found to be very strongly split by spin-orbit (SO) interaction \cite{Minkov14}. Therewith, the SO splitting of the conduction band in the same structures does not reveal itself \cite{Minkov16}.
It is surprising that such SO splitting  is observed in structures both with inverted and normal spectrum despite the fact that at $d<d_c$ and  $d>d_c$ the conduction band is formed from different terms (see Fig.~\ref{F1}). At $d<d_c$, the conduction band is formed from  electron states and states of light hole, while at $d>d_c$,  it is formed from heavy-hole states. Such SO splitting was not described by Byckov-Rashba effect taken  into account within \emph{kP} method.  It was assumed \cite{Minkov16}  that  such a surprising  behavior of SO splitting is a result of the interface inversion asymmetry (IIA) in the HgTe quantum well, which was not taken into account in \emph{kP} calculations in  \cite{Gerchikov90,Zhang01,Novik05,Bernevig06,ZholudevPhD}.

The question arises: how other spin-dependent effects, for example, the Zeeman splitting, depend on the spectrum type -- normal or inverted.  We found only two papers where the Zeeman splitting of electron spectrum was measured in the HgTe quantum wells with the width $d$ which is more or less close to $d_c$ \cite{Molenkamp2014,Zhang04}. In Ref.~\cite{Molenkamp2014},
the Zeeman splitting was determined in a structure with normal spectrum, $d=6.1$~nm, at very large electron density $n=1.4\times10^{12}$~cm$^{-2}$. In Ref.~\cite{Zhang04}, it was determined in a structure with inverted spectrum with  $d=9$~nm that is noticeably larger than $d_c$, at $n=6.59\times  10^{11}$~cm$^{-2}$.

So, up to now a systematic study of the Zeeman splitting and a comparison of it with theoretical calculations are absent.
In this paper, we present the results of  the investigation of the Shubnikov-de Haas (SdH) oscillations in tilt magnetic fields in the  HgTe quantum wells with normal  and   inverted spectra. To find the ratio of the Zeeman splitting to the orbital one, we have used a modified coincidence method which consists in measuring the angle dependence of amplitudes of the SdH oscillations in low magnetic fields.
The simultaneous analysis of this dependence and the shape of oscillations of $\rho_{xx}$ made it possible to determine both the ratio of the Zeeman splitting to the cyclotron one and the anisotropy of $\textsl{g}$-factor ($\textsl{g}_\parallel/\textsl{g}_\perp$) over a wide electron-density range, where $\textsl{g}_\parallel$ and $\textsl{g}_\perp$ are the in-plane and transverse $\textsl{g}$-factor, respectively.

\section{Experiment}
\label{sec:expdet}

Our samples with the HgTe quantum wells  were realized on the basis of
HgTe/Hg$_{1-x}$Cd$_{x}$Te ($x=0.55-0.65$) heterostructures grown by the
molecular beam epitaxy on a GaAs substrate with the (013) surface
orientation \cite{Mikhailov06}.  The samples were mesa etched into standard Hall bars of
$0.5$~mm  width and the distance between the potential probes was
$0.5$~mm. To change and control the carrier density in the quantum
well, the field-effect transistors were fabricated with parylene as an
insulator and aluminium as a gate electrode. For each heterostructure,
several samples were fabricated and studied. The Zeeman splitting of the conduction band  has been obtained  from measurements of the SdH effect in a tilted magnetic field, i.e. we used the so-called coincidence method.
This method is based on the fact that the spin splitting, $\textsl{g}\mu_BB$, depends on the total magnetic field ($B$) whereas the orbital splitting of  the Landau levels (LLs) in 2D systems, $\hbar\omega_c$,   is proportional to the component of the magnetic field which is perpendicular to the 2D plane ($B_\perp$): $\hbar\omega_c=(e\hbar/m_e)B_\perp=(e\hbar/m_e)bB$, where $b\equiv B_\perp/B$.

\begin{table}
\caption{The parameters of  heterostructures under study}
\label{tab1}
\begin{ruledtabular}
\begin{tabular}{ccccc}
number & structure & $d$ (nm) & type & $p,n(V_g=0)$ (cm$^{-2}$)\\
\colrule
  1& 1520 & 4.6   & $n$   &$1.5\times 10^{11}$   \\
  2& 1122 & 5.6    & $p$    & $1.3\times 10^{11}$  \\
  3& 1023 & 6.5     & $p$   & $1.0\times 10^{10}$  \\
  4& H725 & 8.3     & $p$   & $5.0\times 10^{10}$  \\
  5& HT71 & 9.5     & $p$   & $6.0\times 10^{10}$  \\
\end{tabular}
\end{ruledtabular}

\end{table}

Thus, the ratio of the Zeeman splitting to the orbital one,  $\textsl{g}\mu_BB/\hbar\omega_c$, will change upon  varying the  tilt angle as $X(b)=\textsl{g}\mu_B m_e/(e\hbar B\, b)$.
It is clear that there are particular angles  $b_c$ when  $X(b_c) =1/2,1,3/2..$.  At integer $X(b_c)$ values, the energies of the LLs with different numbers and opposite  spin coincide with each other
and the distances between the pairs of such degenerate LLs are equal to  $\hbar\omega_c$. When $X(b_c)$ is half-integer, the energy distances between nearest LLs are twice as low, $0.5\hbar\omega_c$. As a result, the oscillation periods will differ twice for these cases. Knowing values of   $b_c$, one can find the ratio $X(1)\equiv X$. For example, when $X(b_c)=1/2$, $X=b_c/2$.

In this paper, we used the modified coincidence method \cite{Fang1968,Stud05,Kurganova11}.
We have measured the oscillations at low magnetic field when: (i) $B_\perp$  is significantly less than the field of the onset of the quantum Hall effect (QHE); (ii) amplitude of the oscillation is small so that oscillations of the Fermi energy are negligible; (iii) the  SdH oscillations are spin-unsplit. In this case, the study of the angular dependence of the oscillation amplitude $A(b)$ at a given $B_\perp$ value  within the entire range of angles (rather than the determination of critical angles) allows one not only  to determine the ratio of the Zeeman splitting to the orbital one, but  estimate the $\textsl{g}$-factor anisotropy.

To find analytic expression for the tilt-angle dependence of the oscillation amplitude $A(b)$, it is convenient to represent the oscillations as the sum of the contributions from the two series of the Landau spin sublevels.
At low magnetic field, when the SdH oscillations are unsplit, the main contribution to the oscillations of $\rho_{xx}$ comes from the first harmonic,
\footnote{An exception is the case when  $\Delta E=0.5\hbar\omega_c$, and the amplitude of the first harmonic is zero.}
and the well known Lifshits-Kosevitch (LK) formula  for the SdH oscillations  reduces to the following expression
\begin{eqnarray}
\Delta\rho_{xx}(B,b)&=&\rho_{xx}(B,b)-\rho_{xx}(0) \nonumber \\
&=&a_\uparrow \cos\left[2\pi\left(\frac{E_F}{\hbar\omega _c}
+\frac{1}{2}+\frac{X(b)}{2}\right)\right] \nonumber \\
&+&a_\downarrow \cos\left[2\pi\left(\frac{E_F}{\hbar\omega _c}+\frac{1}{2}-\frac{X(b)}{2}\right)\right].
\label{eq1}
\end{eqnarray}
Here, the factors  $a_\uparrow$ and  $a_\downarrow$ depend on the Dingle factor,  temperature, and magnetic field. When $a_\uparrow =a_\downarrow=a$, Eq.~(\ref{eq1}) is
\begin{equation}
\Delta\rho_{xx}(B,b)=
2a \cos{\left[\pi X(b)\right]}\cos{\left[2\pi\left(\frac{E_F}{\hbar\omega _c}+\frac{1}{2}\right)\right]}.
\label{eq2}
\end{equation}
Thus, over  this magnetic field range, the values of $B_\perp$ corresponding to the extremes of $\rho_{xx}$ should not depend on the tilt angle,   while the amplitudes of oscillations $A(b)=2a\cos\left[\pi X(b)\right] $ should periodically change  with $b$ and the angular dependence of the relative amplitude is

\begin{equation}
\frac{A(b)}{A(1)} = \frac{\cos[\pi X(b)]}{\cos[\pi X(1)]}.
\label{eq3}
\end{equation}

\section{Results and discussion}
\label{sec:char}

\begin{figure}
\includegraphics[width=\linewidth,clip=true]{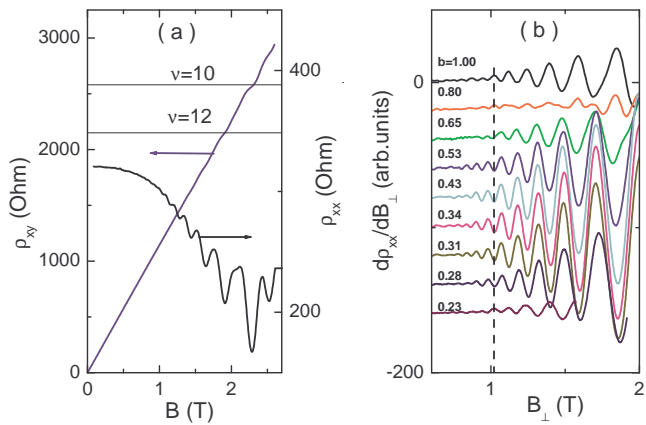}
\caption{(Color online) (a) The magnetic field dependences of $\rho_{xx}$ and $\rho_{xy}$ at normal magnetic field in structure 1520 at $n=5.45\times  10^{11}$~cm$^{-2}$. $T=4.2$~K. (b) The  dependences of $d\rho_{xx}/dB_\perp$ on $B_\perp$ for different $b$. }
\label{F2}
\end{figure}

Let us begin our analysis with the results obtained in the structures with a normal spectrum ($d<d_c$) (see Table~\ref{tab1}). As an example,  consider the results for the structure 1520. Before discussing  the oscillations in the tilt magnetic field, it is necessary to estimate the magnetic field range   where Eq.~(\ref{eq1}) is valid for this structure.
To this end, let us inspect the  magnetic field dependences of
$\rho_{xx}$ and $\rho_{xy}$ in the normal field, which are presented in Fig.~\ref{F2}(a) for the electron density $n=5.45\times 10^{11}$~cm$^{-2}$. It is seen that at $B<1.5$~T, the amplitude of oscillations of $\rho_{xx}$ is $10$ percent less, and the steps in $\rho_{xy}(B)$ (with even filling factors $\nu$) appear only at $B>1.5$~T; therefore one can neglect the oscillations of the Fermi energy within this range of $B$. The electron density  found from the period of oscillations under assumption that the Landau levels are two-fold degenerate, coincides with the Hall density \footnote{The electron density for all structures investigated linearly depends on the gate voltage $V_g$, $n(V_g)= n(0)+\alpha V_g$ and  $\alpha$,  within experimental error, coincides with $C/(S |e|)$, where $C$ is the capacitance measured in the same sample; $S$ is the gate area.}. Thus, at $B<1.5$~T,  the conditions of applicability of  Eq.~(\ref{eq1}) are met.

Now let us inspect the SdH oscillations in the tilt magnetic field. To remove the monotonic part we plotted in   Fig.~\ref{F2}(b)  the $d\rho_{xx}/dB_\perp$ versus $B_\perp$ dependences,  measured  at different tilt angles. To make it easier to trace  the position of oscillations at different angles, we  mark  the position of one of the maxima  $d\rho_{xx}/dB$  at the normal field $B=1.02$~T by a dashed line.  It is clearly seen that  the positions of extremes of $d\rho_{xx}/dB_\perp$  do not change with tilt angle but the maxima are transformed to the minima at $b\simeq0.65$, and upon further rotation they are transformed to the maxima again at $b\simeq0.23$. Note, a noticeable  difference  was not observed when the parallel component of the field was along or perpendicular to the current.

\begin{figure}
\includegraphics[width=0.75\linewidth,clip=true]{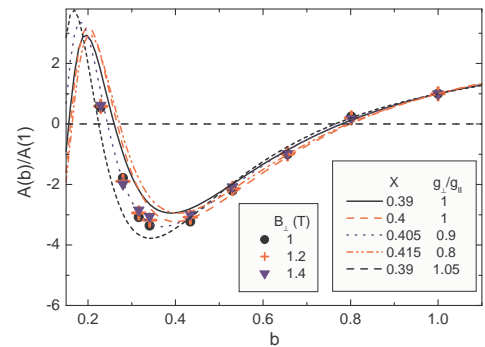}
\caption{(Color online) The  relative oscillations amplitude for the structure H1520 plotted against  $b$ at $n=5.45\times 10^{11}$~cm$^{-2}$. Points are experimental data found at $B_\perp=1.0,\, 1.2$ and $1.4$~T. The inversion of the amplitude sign corresponds to   the change of the oscillation phase by $\pi$. Solid and dash lines  are the dependences  Eq.(\ref{eq3})  with $X=0.39$ and $0.4$, respectively. Other lines are the calculated dependences  with taking into account $\textsl{g}$-factor anisotropy.
% with $\textsl{g}_\perp/\textsl{g}_\parallel=0.9$ and $X=0.405$.
}\label{F3}
\end{figure}

For the quantitative analysis, the amplitude of oscillations  $A(B_\perp,b)$ at a given $b$ was found by fitting of the oscillatory part of $A(B_\perp,b)$ versus $B_\perp$  curves to the oscillating function corresponding to the electron density  $n=1/eR_H$ measured in the normal field, with the amplitude $A(B)=k_0 \exp(k_1 B)$.
The relative amplitudes of the oscillations found in this way
$A(b)/A(1)$ as a function of $b$ are plotted  for some values of $B_\perp$ in Fig.~\ref{F3}. The inversion of the amplitude sign corresponds to the change in the oscillations phase by $\pi$. Note, that $A(b)/A(1)$ does not depend practically on $B_\perp$ when the magnetic field is sufficiently small so that Eq.~(\ref{eq1}) is valid. In the same figure we have shown the dependences which are given by Eq.~(\ref{eq3}) for some values of the ratio $X(b)$=$\textsl{g}\mu_B B/(e \hbar B_\perp/ m_e)$. One can see that Eq.~(\ref{eq3}) well describes the experimental data with $X=0.39\pm0.01$. Some deviation at $b<0.3$ can result from the  $\textsl{g}$-factor anisotropy. Indeed, taking this possibility into account in simplest form
\begin{equation}
\textsl{g}(b)=\sqrt{\textsl{g}_\perp^2 b^2+\textsl{g}_\parallel^2 (1-b^2)}
\label{eq4}
\end{equation}
with $\textsl{g}_\perp/\textsl{g}_\parallel=0.9$  we obtain the exact coincidence over all range of tilt angles with $X=0.405$ (see dot line in Fig.~\ref{F3}). In Fig.~\ref{F3}, we have plotted also the angular dependences of $A(b)/A(1)$  with close pairs of parameters. This makes it possible to assess how uniquely these parameters are  determined.
Note that the value of $X<0.5$ is consistent with the fact that the onset of QHE is observed with  even numbers [see Fig.~\ref{F2}(a)].

\begin{figure}
\includegraphics[width=\linewidth,clip=true]{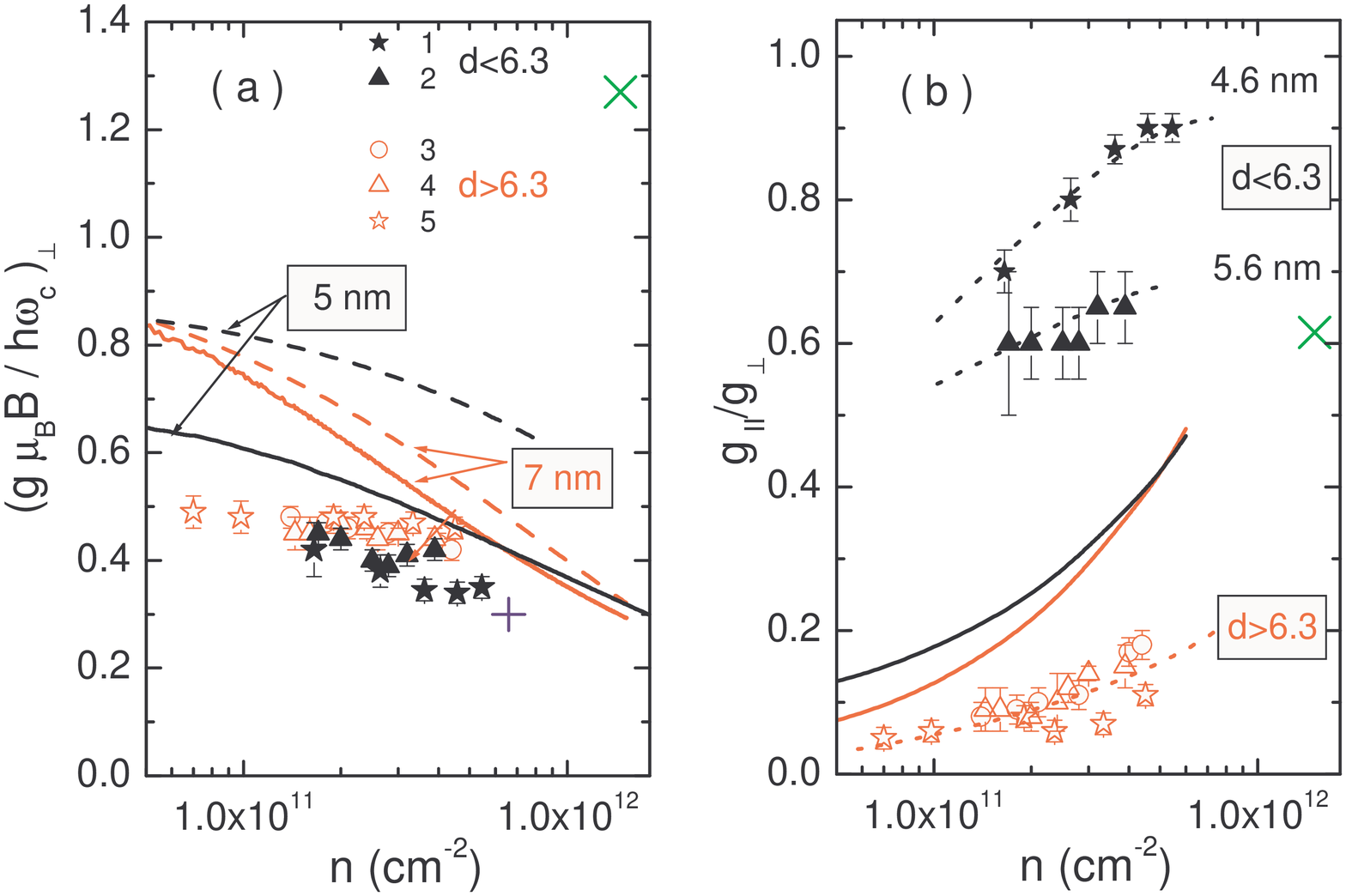}
\caption{(Color online) The  ratio of the Zeeman splitting to the orbital one at normal magnetic field (a) and $\textsl{g}$-factor anisotropy (b) plotted against the electron density. The solid and open symbols are the experimental data for the structures with normal and inverted spectra, respectively. The structures numbers are presented in (a).  The diagonal crosses  are the result of \cite{Zhang04} and straight cross are the result of \cite{Molenkamp2014}. The lines are the calculated dependences for $d=5$~nm and $d=7$~nm (see text).
}\label{F7}
\end{figure}

Such data treatment was carried out for other electron densities and all the results for $\textsl{g} \mu_B B/\hbar\omega_c$ versus the electron density are plotted in Fig.~\ref{F7} together with the results obtained for another structure with $d<d_c$, 1122 (see Table~\ref{tab1}).

Before comparing these results with the theoretical dependences of $\textsl{g} \mu_B B/\hbar\omega_c$ on the electron density, let us consider  the data for structures with inverted spectrum ($d>d_c$).

\begin{figure}
\includegraphics[width=0.9\linewidth,clip=true]{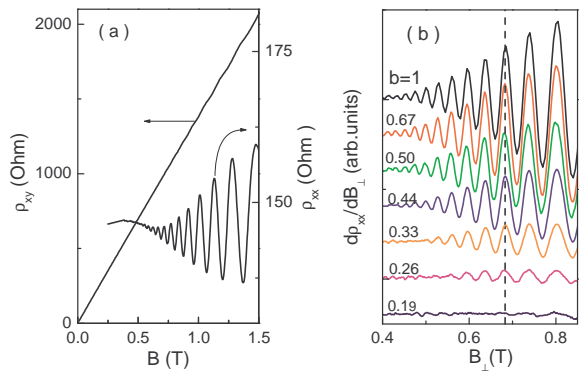}
\caption{(Color online) (a) The magnetic field dependences of $\rho_{xx}$ and $\rho_{xy}$ at normal magnetic field for the structure 1023 ($d<d_c$) at $n=4.6\times  10^{11}$~cm$^{-2}$. $T=4.2$~K. (b) The  dependences of the derivatives $d\rho_{xx}/dB_\perp$ on $B_\perp$ for different $b$.
}
\label{F4}
\end{figure}

As an example, in Fig.~\ref{F4}, we have presented the data for structure 1023 at $n=4.6\times 10^{11}$~cm$^{-2}$. One can see that at $B<0.8$~T, the oscillations of $\rho_{xy}$ are  practically absent and the spin splitting of the oscillations of $\rho_{xx}$ is not observed. So, at $B<0.8$~T, the conditions for applicability of Eq.~(\ref{eq1}) are met. The derivatives $d\rho_{xx}/dB_\perp$, measured at different tilt angles as a function of $B_\perp$ are presented in Fig.~\ref{F4}(b) for different $b$. For clarity, we have plotted  the dashed line at $B_\perp=0.68\text{~T}$ which corresponds to the position of one of maxima. It is clearly seen that with the tilt angle increase, the positions of the extremes of $d\rho_{xx}/dB$  in $B_{\perp}$, similarly to  in structures with $d <d_c$ (see. Fig.~\ref{F2}b), does not change but the amplitude of the oscillations  decreases significantly slower than in the structure with $d<d_c$, and the inversion of the oscillations phase is not observed up to $b=0.19$  [compare Fig.~\ref{F2}(b) and Fig.~\ref{F4}(b)].

\begin{figure}
\includegraphics[width=\linewidth,clip=true]{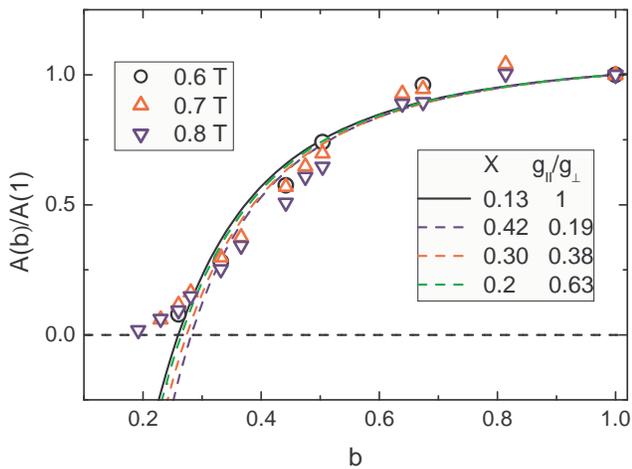}
\caption{(Color online) The  relative oscillation amplitude for  structure 1023 plotted against  $b$. Points are experimental data found at $B_\perp=0.6, 0.7$, and $0.8$~T at the electron density $n=4.6\times 10^{11}$~cm$^{-2}$. The lines are the dependences  Eq.~(\ref{eq3}) with different pairs of the parameters $X$ and  $\textsl{g}_\parallel/\textsl{g}_\perp$ shown in figure.
}\label{F5}
\end{figure}

The dependence of the amplitude of oscillations  on $b$ together with the calculated dependence,  Eq.~(\ref{eq3}),  with the isotropic $\textsl{g}$-factor  is presented in Fig.~\ref{F5}. One can see that this simple dependence describes well the data over $b$ range from $1$ to $(0.25-0.2)$ with $X=0.13$. Note, this value is three-four times as low as that for the structures with normal spectrum (see Fig.~\ref{F3}).  Let us check  how unambiguously the value of $X$ is determined for this case. To this end, we have plotted in Fig.~\ref{F5} the  $A(B_\perp,b)$ versus $b$ dependences which were calculated  using two free parameters, namely $\textsl{g}_\parallel/\textsl{g}_\perp$ and $X$. One can see that the experimental data are equally well described with very different pairs of  $X$ and $\textsl{g}_\parallel/\textsl{g}_\perp$.

\begin{figure}
\includegraphics[width=\linewidth,clip=true]{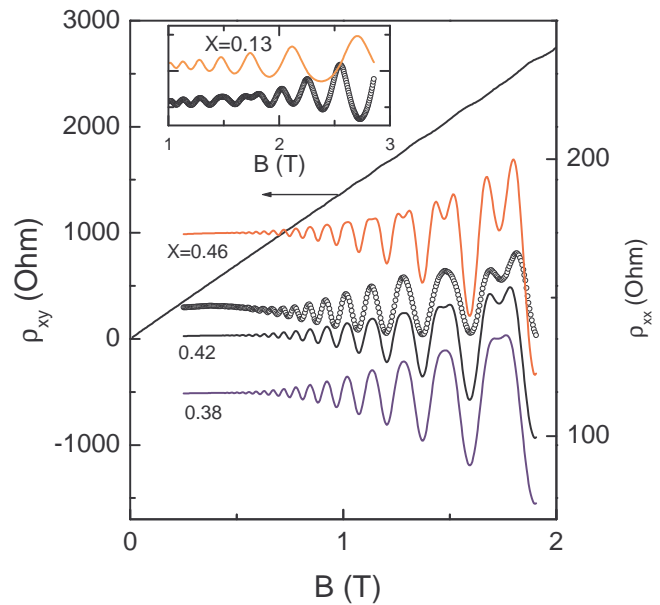}
\caption{(Color online) The experimental dependences of $\rho_{xx}$ and $\rho_{xy}$ for  structure 1023 in the normal magnetic field at $n=4.6\times  10^{11}$~cm$^{-2}$ for magnetic field range  larger than in Fig.~\ref{F4} (points). The solid lines are the calculated $\rho_{xx}$ curves with different values of $X$. These curves are shifted for clarity.  The inset shows comparison of the data with the calculation with $X=0.13$ up to $B=3$~T (see text).
}\label{F6}
\end{figure}

To avoid such a large ambiguity, let us consider oscillations of $\rho_{xx}(B)$ in a  larger magnetic field, where the Zeeman splitting  starts  to be observed but lower than the  onset of QHE. Such experimental dependence in the  normal field is presented in Fig.~\ref{F6} together with the curves calculated using the LK formula with different $X$ values. We assumed the Lorentz  broadening  of LLs with parameter $\Delta_L=4.5$~meV  which was found from the magnetic field dependence of $\rho_{xx}(B)$ at $B<1$~T.
 The inset shows that the calculated curve with $X=0.13$ radically differs from the experimental curve, it does not demonstrate the Zeeman splitting of the oscillations up to $3$~T. The calculated curves  with  $X=0.35 - 0.48$ are significantly closer to the experimental dependence $\rho_{xx}(B)$, therewith the curve with $X=0.42$   practically reproduces the experimental dependence.
Thus, the comparison of the data with the calculated curves  in Fig.~\ref{F5} and Fig.~\ref{F6} gives the possibility  to find  unambiguously  the values  of  $X$ and $\textsl{g}_\parallel/\textsl{g}_\perp$ as  $0.42\pm0.03$ and  $0.19\pm0.02$, respectively.

As seen from Fig.~\ref{F5}, some discrepancy between the $A(b)/A(1)$  data  and calculated curves remains at $b<0.2$.
The reasons for this discrepancy are not clear.  TThey may be: (i) effect of sufficiently  large in-plane component of $B$ because  at $b=0.2$, the  in-plane component of $B$ is about $3-4$ T, which corresponds to the magnetic length $L=(13-14)$~nm so that $L$ became comparable to the  width of the quantum well width $(8-9)$~nm. That can change the energy spectrum noticeably. (ii) a difference in the broadening (or difference in contribution to the oscillations) of different spin sub-levels; (iii) imperfect flatness of the quantum well and so on.
Nevertheless, we believe that the value of $X=0.42\pm 0.02$  corresponds to the  ratio of the Zeeman splitting  to the cyclotron energy in the  normal magnetic field and $\textsl{g}$-factor anisotropy is $\textsl{g}_\parallel/\textsl{g}_\perp=0.19\pm0.02$.

The described above measurements and data treatment were carried out  for all structures from Table~\ref{tab1} over the wide electron density range. All obtained values of $X$ and $\textsl{g}_\parallel/\textsl{g}_\perp$ versus the electron density are  summarized in Fig.~\ref{F7}.

One can see that  the ratio of the Zeeman splitting to the orbital one is close to each other for both types of the structures, with the normal and inverted spectrum.  This ratio decreases slightly  from $X\approx 0.5$ to $\approx 0.4$, as the electron density increases  from $n\approx 1\times 10^{11}$~cm$^{-2}$ to $5\times 10^{11}$~cm$^{-2}$.

The values of the $\textsl{g}$-factor anisotropy in the structures with normal and inverted spectra  differ  significantly [see Fig.~\ref{F7}(b)]. The values of $\textsl{g}_\parallel/\textsl{g}_\perp$ in the structures with the normal spectrum  are in the range of  $0.6-0.9$, while  in the structures with inverted spectrum they are in the range of $0.05 - 0.015$. For both types of the structures the values of $\textsl{g}_\parallel/\textsl{g}_\perp$ increase with increasing electron density.

Let us compare our data with the results of previous studies.
We have found only two articles \cite{Molenkamp2014,Zhang04} where the Zeeman splitting was studied in the structures with $d\approx d_c$ and we  plotted them in Fig.~\ref{F7}. In paper \cite{Zhang04}, the Zeeman splitting was determined for the normal magnetic field only in the structure with $d=9$~nm at the  electron density $n=6.59\times 10^{11}$~cm$^{-2}$. The value of $X$ agrees well with our data (see Fig.~\ref{F7}).  In  Ref.~\cite{Molenkamp2014}, both the  Zeeman splitting and $\textsl{g}$-factor anisotropy were found for the structure with $d=6.1$~nm for the very high electron density, $n=1.46\times 10^{12}$~cm$^{-2}$.
The value of the $\textsl{g}$-factor anisotropy is found to be close to our data for the structure with $d=5.6$~nm [see Fig.~\ref{F7}(b)], while the ratio $\textsl{g}\mu_BB/\hbar\omega_c$ is significantly larger than our data: $1.26$ instead of $0.4-0.5$. Such difference is unclear. One of possible reasons is  role of spin-orbit interaction, which can be large for so high electron density and  was not taken into account in the analysis of the data.

Now let us compare the obtained results with the theoretical ones. To find $X$=$\textsl{g}\mu_B B/(e \hbar B_\perp/ m_e)$,   the positions of the Landau levels have been calculated in framework of the $8$-band \emph{$kP$} model \cite{ZholudevPhD}. Since there are different notations of the Landau levels in various papers, we have numbered the levels in a row, starting from unity for the lowest LL of the conduction band. The Zeeman splitting was found as the energy distance between the levels $n$ and $n+1$ with odd $n$,  while the orbital splitting was found as the distance between the levels $n$ and $n+2$ \footnote{The facts that the onset of QHE is observed with  even numbers and  the amplitude of the oscillations decreases at deviation of the magnetic field from the normal to 2D plane show that the ratio of namely these values is determined from the dependence $A(b)$}. The calculated $\textsl{g} \mu_B B/\hbar\omega_c$ versus $n$ dependences are plotted in Fig.~\ref{F7}(a) by solid lines. It is seen that the experimental values  are slightly ower than the calculated ones  for the structures with the normal and inverted spectrum. It is instructive to compare the results of the calculations performed  in the framework of the  8-band \emph{kP} model with those obtained within the framework of the  Bernevig-Hughes-Zhang (BHZ) model \cite{Bernevig06}, which is often used to analysis   various effects. We have used the parameters of the BHZ model which give the dependence $E(k)$ very close to that calculated in framework of the  $8$-band \emph{kP} model. However, the Zeeman splitting in this case  appears to be $20-30$ percent larger [see dashed lines in Fig.~\ref{F7}(a)].

To compare the data for the $\textsl{g}$-factor anisotropy with the theory, one needs to know the values of $\textsl{g}_\parallel$ together with $\textsl{g}_\perp$ calculated just above.
The dependences  of  $\textsl{g}_\parallel$ on electron density were calculated  using the results of the paper  \cite{Raichev12-1} where the energy spectrum of the HgTe quantum wells in the in-plane magnetic field was studied. The calculated dependences of $\textsl{g}_\parallel/\textsl{g}_\perp$ versus electron density are shown in Fig.~\ref{F7}(b). It is seen that  for both types of the structures, with the normal and inverted spectrum, the theoretical values of $\textsl{g}_\parallel/\textsl{g}_\perp$ are small, they are close to each other, and increase with the electron density increase. The calculated values of $\textsl{g}_\parallel/\textsl{g}_\perp$  significantly differ from the experimental data for both types of the structures. For the structures with the inverted spectrum ($d>d_c$), the experimental values are  to $1.5-2$ times lower. For  the structures with the normal spectrum, the difference is larger and  the experimental data are to $3-4$ times higher than the calculated ones.

To discuss  possible reasons for the discrepancy, let us remind the results of our previous paper \cite{Minkov16}.
We have shown (see Introduction) that for $d\approx d_c$ in nominally symmetric structures the top of the valence band is very strongly split by SO interaction \cite{Minkov14}. Therewith, the SO splitting of the conduction band in the same structures does not reveal itself \cite{Minkov16}. It is surprising that such SO splitting  is observed  in structures with the inverted and normal spectrum despite the fact that at $d<d_c$ and  $d>d_c$ the conduction band is formed from different terms (see Fig.~\ref{F1}). It was assumed in \cite{Minkov16}  that  such surprising  behavior of the SO splitting is a result of the interface inversion asymmetry in the HgTe quantum well, which is not taken into account in \emph{kP} calculations \cite{Gerchikov90,Zhang01,Novik05,Bernevig06,ZholudevPhD}.
We believe that the  disagreement between the experimental data on  $\textsl{g}$-factor anisotropy and calculations is also a result of the interface inversion asymmetry in the HgTe quantum well, which is not taken into account in \emph{kP} calculations.

In summary, the ratio of the Zeeman splitting to the orbital
one and anisotropy of the $\textsl{g}$-factor in the  HgTe quantum
wells both with normal and inverted spectrum have been studied experimentally within a wide electron density range. To obtain two these parameters unambiguously,  we have analyzed both the tilt angle dependence of the SdH oscillations in low magnetic fields and the shape of the oscillations
in moderate magnetic fields. It has been shown
that the ratios of the Zeeman splitting to the orbital one
are close to each other in the structures with normal and
inverted spectra, these ratios decrease when the electron
density increases and they are quite close to the values calculated
within the \emph{kP} method. In contrast, the anisotropy
of $\textsl{g}$-factor in the structures with the normal and inverted spectrum is strongly different and for both cases differ significantly
from the calculated ones. We believe that such
disagreement with the calculations is a result of the interface inversion
asymmetry in the HgTe quantum well, which is not
taken into account in the \emph{kP} calculations.

\acknowledgements

We are grateful to  S.~Studenikin, V.~Aleshkin, A.\,V.~Germanenko and O.\,E. Raichev   for useful discussions and M.~Zholudev for calculations of LLs in framework of 8-band $kP$ model.
The work has been supported in part by the Russian Foundation for Basic
Research (Grants No. 16-02-00516 and No. 15-02-02072) and by  Act 211 Government of the Russian Federation, agreement No. 02.A03.21.0006. A.V.G. and O.E.R. gratefully acknowledge financial support
from the Ministry of Education and Science of the Russian
Federation under Projects No. 3.571.2014/K and No. 2457.

%\bibliography{QuantumCorrections1}
%\bibliography{QuantumCorrections}

\end{document}